\documentclass[aps,pra,reprint,superscriptaddress,showkeys]{revtex4-1}
\usepackage[english]{babel}
\usepackage{graphicx}
\usepackage{siunitx}
\usepackage{amsmath}

\newcommand{\figref}[1]{Fig.~\ref{#1}}

\begin{document}
	
	\title{Substrate-Induced Chirality in an Individual Nanostructure}
	
	\author{Sergey Nechayev} 
	\thanks{These authors contributed equally}
	\affiliation{Max Planck Institute for the Science of Light, Staudtstr. 2, D-91058 Erlangen, Germany}
	\affiliation{Institute of Optics, Information and Photonics, University Erlangen-Nuremberg, Staudtstr. 7/B2, D-91058 Erlangen, Germany}
	
	\author{Ren\'{e} Barczyk}
	\thanks{These authors contributed equally}
	\affiliation{Max Planck Institute for the Science of Light, Staudtstr. 2, D-91058 Erlangen, Germany}
	\affiliation{Institute of Optics, Information and Photonics, University Erlangen-Nuremberg, Staudtstr. 7/B2, D-91058 Erlangen, Germany}

	\author{Uwe Mick}
	\affiliation{Max Planck Institute for the Science of Light, Staudtstr. 2, D-91058 Erlangen, Germany}
	\affiliation{Institute of Optics, Information and Photonics, University Erlangen-Nuremberg, Staudtstr. 7/B2, D-91058 Erlangen, Germany}
	
%
%
	\author{Peter Banzer}
	\email{peter.banzer@mpl.mpg.de}
	\homepage{http://www.mpl.mpg.de/}
	\affiliation{Max Planck Institute for the Science of Light, Staudtstr. 2, D-91058 Erlangen, Germany}
	\affiliation{Institute of Optics, Information and Photonics, University Erlangen-Nuremberg, Staudtstr. 7/B2, D-91058 Erlangen, Germany}
	\date{\today}

	\begin{abstract}
		We experimentally investigate the chiral optical response of an individual nanostructure consisting of three equally sized spherical nanoparticles made of different materials and arranged in \ang{90} bent geometry. Placing the nanostructure on a substrate converts its morphology from achiral to chiral. Chirality leads to pronounced differential extinction, i.e., circular dichroism and optical rotation, or equivalently, circular birefringence, which would be strictly forbidden in the absence of a substrate or heterogeneity. This first experimental observation of the substrate-induced break of symmetry in an individual heterogeneous nanostructure sheds new light on chiral light-matter interactions at substrate-nanostructure interfaces.
	\end{abstract}
	
	\maketitle

%
\section{Introduction} 

\begin{figure}[!t]
		\centering 
		\includegraphics[width=0.48\textwidth]{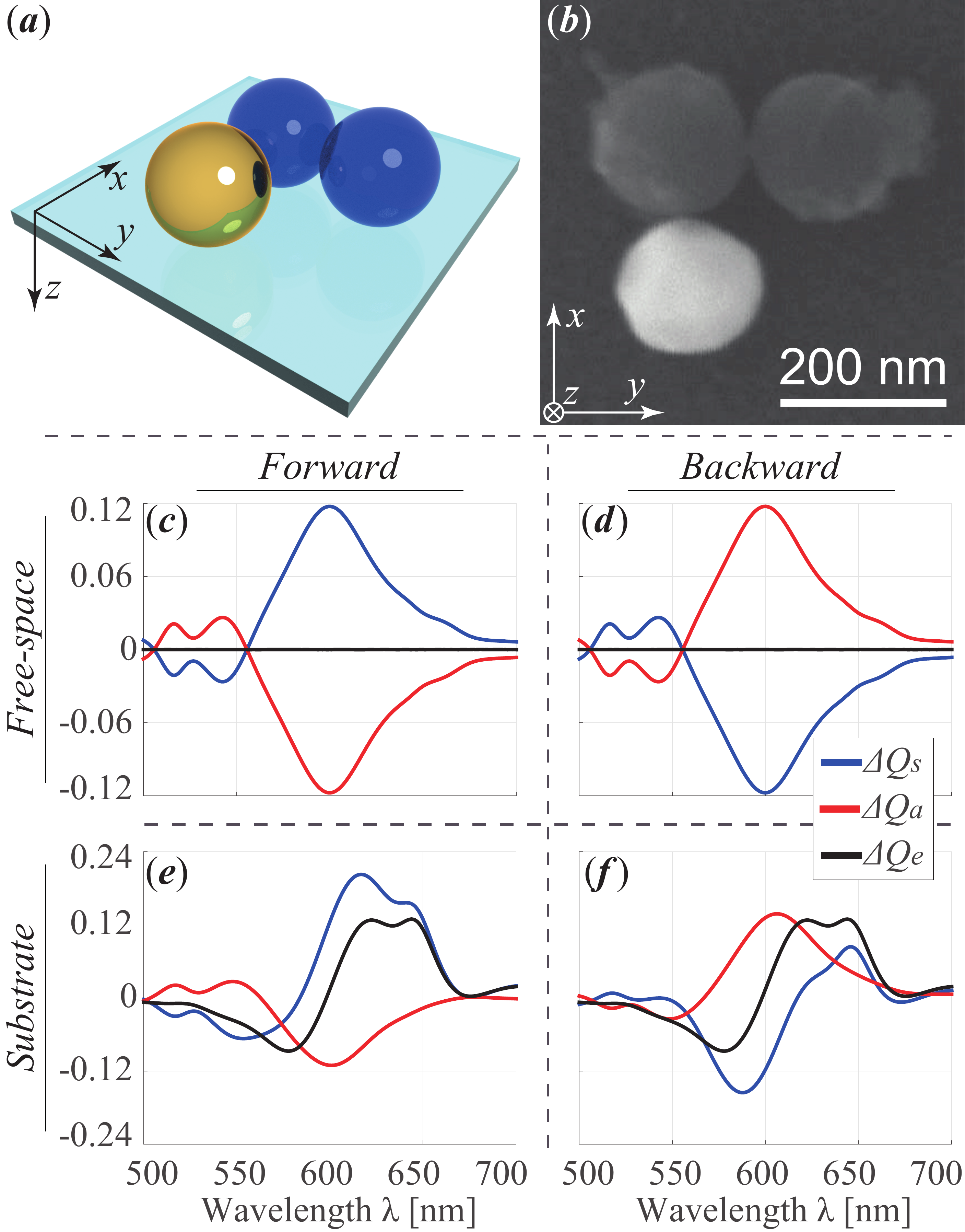} 
		\caption{$(\boldsymbol{a})$ Schematic illustration and $(\boldsymbol{b})$ scanning electron microscope image of the investigated nanotrimer. The sample is composed of three nanospheres with radii $r=90\,\mathrm{nm}$, arranged in \ang{90} bent geometry and positioned on a glass substrate. The two upper nanoparticles in $(\boldsymbol{b})$  are made of silicon (Si), while the third one is made of gold (Au). The estimated gap between neighboring particles is $2\,\mathrm{nm}$. $(\boldsymbol{c})$ Normalized differential scattering ($\Delta Q_s$), absorption ($\Delta Q_a$) and extinction ($\Delta Q_e$) cross-sections for the nanotrimer in free-space, illuminated with left- and right-hand circularly polarized plane-waves along the positive direction of the $z$-axis ($+ \hat {\mathbf{z}}$). $(\boldsymbol{e})$ -- same as $(\boldsymbol{c})$, for the nanotrimer on a glass substrate with a refractive index of $n=1.52$. $(\boldsymbol{d})$ and $(\boldsymbol{f})$ -- same as $(\boldsymbol{c})$ and $(\boldsymbol{e})$, respectively, for illumination from the opposite side ($-\hat {\mathbf{z}}$).}
		\label{fig:fig1}
	\end{figure}

	Lord Kelvin defines chirality as the property of a geometrical figure who's ``image in a plane mirror, ideally realized, cannot be brought to coincide with itself''~\cite{kelvin_baltimore_1904}. Chiral molecules and nanostructures exhibit circular anisotropies~\cite{snatzke1968circular,tang_optical_2010} --- left- and right-hand circular polarizations (LCP and RCP) experience different real and imaginary parts of the refractive index. The former is the origin of circular birefringence (CB), while the latter results in circular dichroism (CD)~\cite{lindell1994electromagnetic,barron2009molecular}. Importantly, circular anisotropies are invariant with respect to reversal of the propagation direction of circularly polarized light (CPL)~\cite{true_false,arteaga_o._mueller_2010,kuwata-gonokami_giant_2005,Maslovski_2009,arteaga_complete_2014,arteaga_relation_2016} and their proper measurement requires some caution.\\
	For instance, CD is usually defined as a measure of differential extinction of CPL and therefore necessitates polarization analysis of the transmitted light~\cite{arteaga_o._mueller_2010}, unless it is observed as an average value of monodisperse solutions~\cite{barron2009molecular}. In the presence of linear anisotropies, which are typical for chiral structures~\cite{govorov_nanocrystals_2012}, the difference in intensity of the transmitted light ($\Delta T$) for LCP and RCP illumination along a certain direction does not necessarily represent CD as defined above, but rather a combination of circular and linear anisotropies~\cite{oriented_mol_1987,arteaga_o._mueller_2010,arteaga_complete_2014,arteaga_relation_2016}. Contrary to CD and CB, linear anisotropies invert their sign upon wavevector reversal~\cite{arteaga_o._mueller_2010,arteaga_complete_2014,arteaga_relation_2016} and the average value of $\left\langle \Delta T\right\rangle$ for backward and forward illumination represents CD~\cite{arteaga_o._mueller_2010,arteaga_complete_2014,arteaga_relation_2016,hentschel_three-dimensional_2012}.\\
	The direction of illumination itself plays an important role for the optical response of anisotropic objects~\cite{oriented_mol_1987,govorov_nanocrystals_2012,Korger:13_polarizer}. Even achiral planar structures may show pronounced tunable CD and CB under oblique illumination, which is typically referred to as extrinsic or pseudo-chirality~\cite{verbiest_optical_1996,Verbiest:98,plum_metamaterials:_2009,singh_highly_2010,sersic_ubiquity_2012,nanotubes_2014,ext_chir_2014,leon_strong_2015,dimer}. Consequently, a measurement of CD and CB along a fixed direction of illumination does not necessarily indicate structural chirality.\\ 
	A practically interesting case are quasi-planar nanostructures (QPNs) with broken in-plane reflection symmetry (or asymmetric QPNs), such as a flat spiral of finite thickness or an asymmetric planar arrangement of arbitrarily sized spheres. Since the sense of twist of a flat spiral inverts with the reversal of the direction of observation and circular anisotropies are invariant under wavevector reversal~\cite{true_false,arteaga_o._mueller_2010,kuwata-gonokami_giant_2005,Maslovski_2009,arteaga_complete_2014,arteaga_relation_2016}, QPNs can not show any CD or CB when illuminated normally to their inherent plane of mirror symmetry. However, QPNs may show a strong chiroptical response in differential transmission ($\Delta T$), differential absorption ($\Delta A$), differential scattering and asymmetric polarization conversion of CPL~\cite{papakostas2003optical,Reichelt2006,fedotov2006asymmetric,fedotov2007asymmetric,Husu_2008,Valev_rachet_2009,Plum_planar_meta_2009,Plum_2010,Zhukovsky:11,eftekhari2012strong,arteaga_complete_2014,arteaga_relation_2016,banzer_chiral_2016,single_particle_cd_2018}. All aforementioned differential measures must invert their sign with the reversal of the illumination direction, if the QPNs are embedded in a homogeneous background. This fundamental difference between the illumination direction-dependent $\Delta T \backslash \Delta A$ and strictly forbidden circular anisotropies attracted significant attention and was discussed in the context of optical reciprocity~\cite{kuwata-gonokami_giant_2005,fedotov2006asymmetric,fedotov2007asymmetric,Plum_planar_meta_2009,Plum_2010,arteaga_complete_2014,true_false,bai2007optical,Maslovski_2009,arteaga_relation_2016,hopkins_circular_2016}.\\
	However, for experimental investigation, QPNs are commonly positioned on a substrate, which breaks the forward-backward symmetry for normally incident CPL. Furthermore, a substrate converts the morphology of the system from achiral to chiral. Substrate-induced emergence of CD and CB that are invariant under wavevector reversal has been experimentally confirmed in arrays of asymmetric QPNs~\cite{kuwata-gonokami_giant_2005,arteaga_relation_2016} and in asymmetric arrays of nanoholes~\cite{arteaga_complete_2014}. At the same time, non-zero $\Delta A$ has been demonstrated for individual nanohelices~\cite{Wozniak:18} and single asymmetric QPNs under normally incident CPL~\cite{banzer_chiral_2016,single_particle_cd_2018}. Differential scattering of CPL was shown for symmetric QPNs under oblique illumination~\cite{ext_chir_2014} and a variety of individual nanostructures under normal incidence~\cite{lin_all-optical_2019}. However, to the best of our knowledge, substrate-induced emergence of differential extinction (CD) and CB in an individual QPN under normal incidence and, hence, its conversion into chiral morphology, have not been experimentally investigated to date.\\
	Here, we apply back-focal plane (BFP) or $\mathbf{k}$-space Mueller matrix spectropolarimetry~\cite{arteaga_o._mueller_2010,osorio_k-space_2015,arteaga_complete_2014,koendering_polarimetry_2015,arteaga_relation_2016} to investigate the emergence of substrate-induced chirality in an individual asymmetric QPN. The QPN consists of three nanospheres of radii $r=90\,\mathrm{nm}$ arranged in \ang{90} bent geometry~\cite{banzer_chiral_2016} with estimated gaps of $2\,\mathrm{nm}$ between neighboring particles (\figref{fig:fig1}$\boldsymbol{a}$ and \ref{fig:fig1}$\boldsymbol{b}$), which are positioned on a glass substrate using a pick-and-place procedure~\cite{bartenwerfer_towards_2011,mick_afm-based_2014}. The in-plane reflection symmetry of the nanotrimer is broken by its heterogeneous material composition~\cite{banzer_chiral_2016} --- the two upper nanoparticles in \figref{fig:fig1}$\boldsymbol{b}$ are made of silicon (Si)~\cite{edited_by_edward_d._palik_handbook_1985}, while the third one is made of gold (Au)~\cite{johnson_optical_1972}. The glass substrate breaks the forward-backward symmetry under normal incidence and renders the whole system structurally chiral. We experimentally reconstruct the emerging CD and CB spectra, which would be strictly forbidden in the absence of a substrate or heterogeneity. 
	%

\section{Results} 

	\begin{figure*}[!t]
		\centering 
		\includegraphics[width=\textwidth]{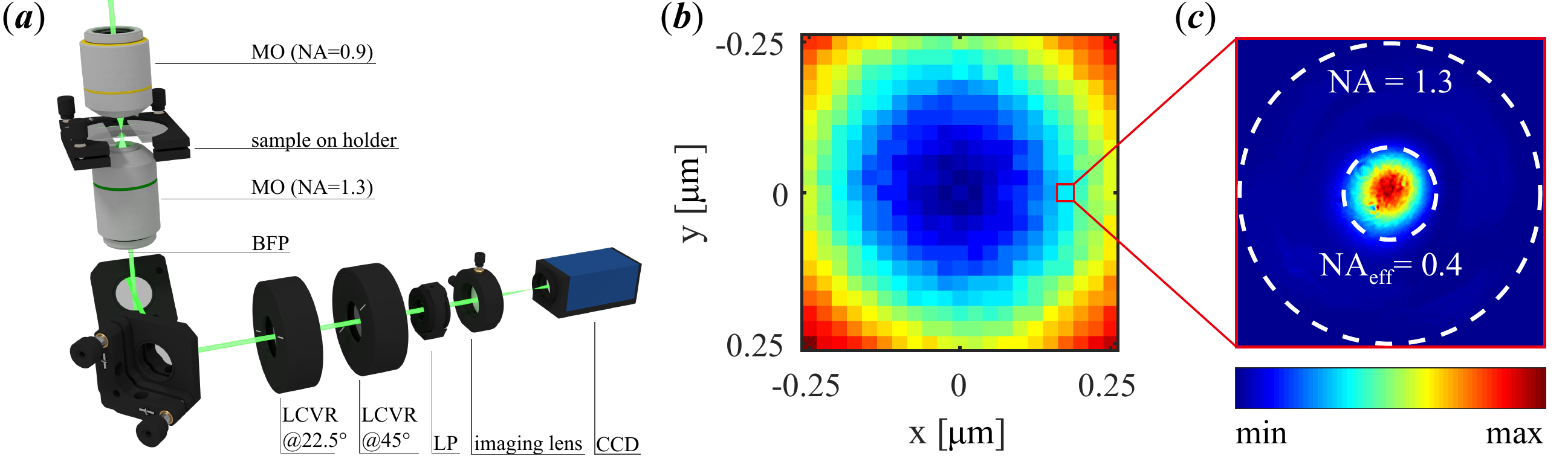} 
		\caption{$(\boldsymbol{a})$ Simplified experimental scheme. The nanotrimer on a glass substrate is mounted on a sample holder, which is attached to a three-dimensional (3D) piezo actuator for precise positioning in the focus. The incident Gaussian beam partly fills the aperture of the upper microscope objective (MO) with numerical aperture (NA) of $\text{NA}=0.9$ and is focused onto the nanotrimer with an effective NA of $\text{NA}_{\text{eff}}=0.4$. The transmitted and scattered light is collected and collimated by the second MO with $\text{NA}=1.3$ and passes two liquid crystal variable retarders (LCVRs, slow axes at \ang{22.5} and \ang{45}) and a linear polarizer (LP) for Stokes analysis via projection onto different polarization states. The polarization-filtered intensity distribution in the back-focal plane (BFP) of the lower MO is imaged onto a charge-coupled device (CCD) camera by a lens. $(\boldsymbol{b})$ Raster scan image of the total transmitted intensity, obtained by moving the nanotrimer sample in a $500\times500\,\text{nm}^2$ region around the focus. We average the results of the polarimetric analysis on a $10\times10$ grid of positions ($250\times250\,\text{nm}^2$) around the estimated center of the sample, indicated by minimum transmission. $(\boldsymbol{c})$ Typical BFP image of the lower MO, recorded for a specific polarization projection and position in the scan grid. White rings indicate the NA of the incident ($\text{NA}_{\text{eff}}=0.4$) and collected ($\text{NA}=1.3$) light. We perform our polarization analysis in the central angular region, where the incident and the scattered light interfere.}
		\label{fig:fig2abc}
	\end{figure*}
\begin{figure}[!tb]
		\centering 
		\includegraphics[width=\columnwidth]{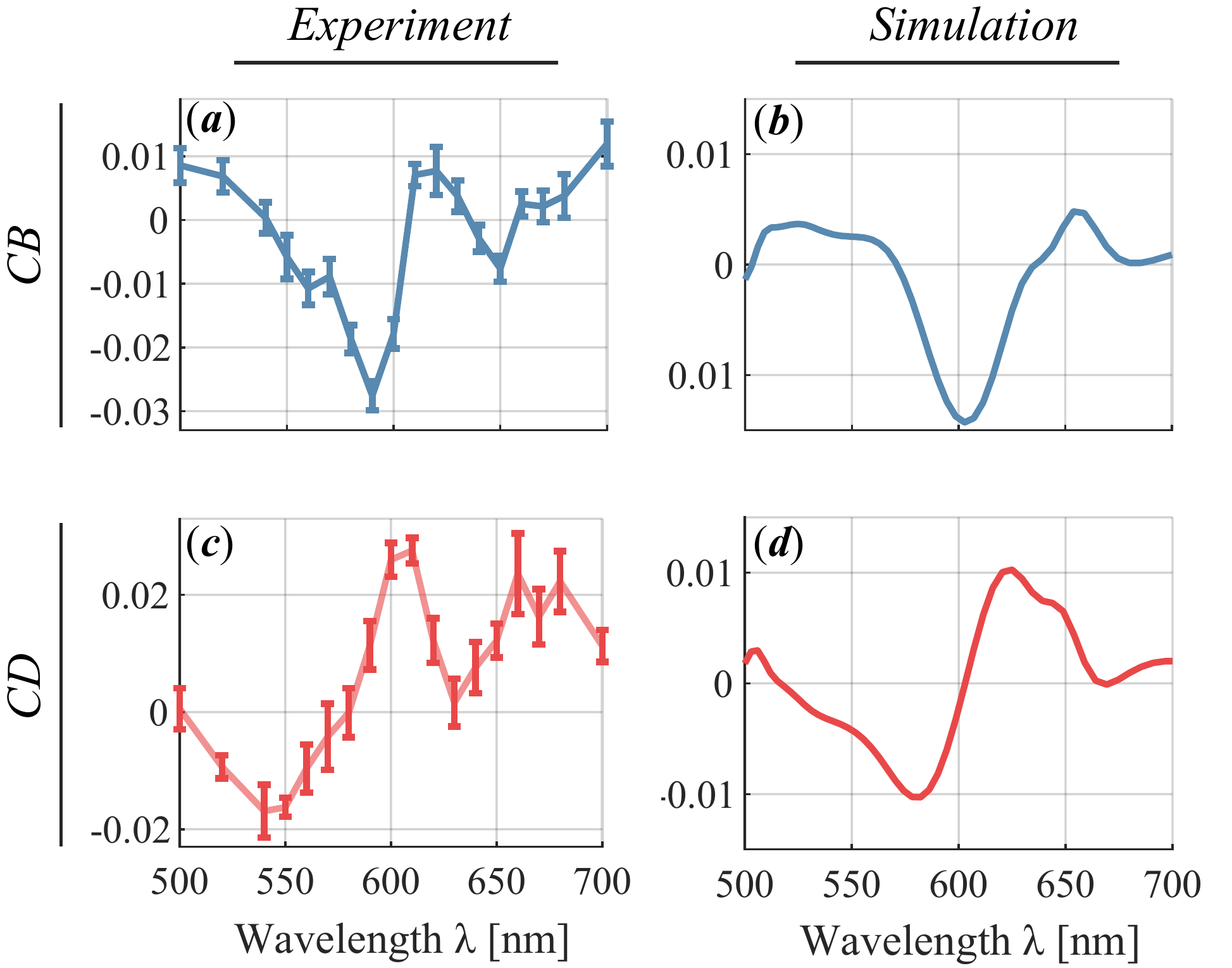} 
		\caption{$(\boldsymbol{a},\,\boldsymbol{c})$ Experimentally measured and $(\boldsymbol{b},\,\boldsymbol{d})$ numerically calculated spectra of $(\boldsymbol{a},\,\boldsymbol{b})$ circular birefringence (CB) and $(\boldsymbol{c},\,\boldsymbol{d})$ circular dichroism (CD) of the investigated nanotrimer sample. The error-bars in $(\boldsymbol{a},\,\boldsymbol{c})$ denote the standard-mean deviation for the spatial average performed over the scan grid.}
		\label{fig:fig3abcd}
	\end{figure}

 
	We start by theoretically investigating the scattering properties of the nanosphere assembly. The sample exhibits a rich spectral behavior with several resonances, residing in the excitation and interaction of various electric and magnetic multipoles in the nanospheres constituting the nanotrimer \cite{dimer,banzer_chiral_2016}. We employ a coupled-dipole model (CDM) to calculate the scattering ($\sigma_s$), absorption ($\sigma_a$) and extinction ($\sigma_e$) cross-sections for normally incident plane-wave CPL illumination~\cite{mulholland_light_1994,albella_low-loss_2013}. In the CDM, each of the nanoparticles is modeled as a point-dipole, whose electric- and magnetic-dipole polarizability is obtained from Mie theory in free-space~\cite{huffman_1983}. Each of these dipoles reacts to the incident field, to the field of the other dipoles and its own reflected field~\cite{Substrates_Matter_halas_2009,miroshnichenko_substrate-induced_2015,mulholland_light_1994,albella_low-loss_2013}. In \figref{fig:fig1}$\boldsymbol{c}$, we plot the differential cross-sections normalized to the geometric cross-section ($\Delta Q_j \equiv \left[\sigma_j^{\text{LCP}}-\sigma_j^{\text{RCP}} \right]/ \left[ 3 \pi r^2 \right] $) for the nanotrimer in free-space, illuminated with LCP and RCP along the positive direction of the $z$-axis ($+ \hat {\mathbf{z}}$). \figref{fig:fig1}$\boldsymbol{c}$ shows that the differential scattering $\Delta Q_s$ has the same magnitude and opposite sign as the differential absorption $\Delta Q_a$, resulting in zero differential extinction $\Delta Q_e=\Delta Q_s+\Delta Q_a=0$. With the reversal of the direction of illumination ($ -\hat {\mathbf{z}}$), as shown in \figref{fig:fig1}$\boldsymbol{d}$, $\Delta Q_s$ and $\Delta Q_a$ just interchange their amplitudes, preserving $\Delta Q_e=0$. \figref{fig:fig1}$\boldsymbol{e}$ shows the differential cross-sections in the presence of a glass substrate ($n=1.52$) for incidence from the air side ($+ \hat {\mathbf{z}}$). $\Delta Q_s$ and $\Delta Q_a$  no longer balance each other, resulting in a non-zero $\Delta Q_e$. Here, reversal of the direction of illumination ($- \hat {\mathbf{z}}$), as shown in \figref{fig:fig1}$\boldsymbol{f}$, strongly affects the spectra of $\Delta Q_s$ and $\Delta Q_a$. Nevertheless, the same $\Delta Q_e \neq 0$ is retained. The CDM~\cite{mulholland_light_1994,albella_low-loss_2013} allows us to understand the origin of non-zero differential extinction of normally incident CPL in the presence of a substrate. Only when the field radiated by each of the nanospheres reflects from the substrate~\cite{Sipe:87,miroshnichenko_substrate-induced_2015,Substrates_Matter_halas_2009} and re-excites the nanospheres, we observe $\Delta Q_e \neq 0$. However, the CDM assumes only \textit{point-dipoles}, located at the respective centers of the nanoparticles. Therefore, the CDM cannot account for strong near-field enhancement in the gaps between the actual nanoparticles, which significantly contributes to the scattering, absorption and extinction spectra. This field enhancement originates from nonradiative higher-order modes~\cite{albella_low-loss_2013} and therefore requires full-wave simulations. For this reason, from this point onwards we employ finite-difference time-domain (FDTD) simulations~\cite{noauthor_fdtd_nodate} for a comparison with the experimental results.\\
	A simplified sketch of the experimental scheme is depicted in \figref{fig:fig2abc}$\boldsymbol{a}$. It consists of two microscope objectives (MOs) in confocal configuration for focusing and collimation of the incident light \cite{banzer_experimental_2010,banzer_chiral_2016}. A three-dimensional (3D) piezo actuator positions the mounted sample in the focal plane. The incident light only partly fills the aperture of the upper MO with numerical aperture (NA) of $\text{NA}=0.9$, such that the nanotrimer is effectively illuminated by a weakly focused Gaussian beam with $\text{NA}_{\text{eff}}=0.4$ from the air side ($+\hat {\mathbf{z}}$). The transmitted and scattered light is collected by a second oil immersion MO with $\text{NA}=1.3$. The beam then passes two liquid crystal variable retarders (LCVRs) (slow axes at \ang{22.5} and \ang{45}) and a linear polarizer (LP) for projection onto different polarization states~\cite{bueno_polarimetry_2000}. Finally, a lens images the polarization-filtered intensity distribution in the lower objective's BFP onto a charge-coupled device (CCD) camera.\\
	The position of the sample relative to the excitation beam is crucial for properly performing the spectropolarimetry of an individual nanostructure. \figref{fig:fig2abc}$\boldsymbol{b}$ shows a raster scan image of the total transmitted intensity, obtained by moving the nanotrimer sample in a $500\times500\,\text{nm}^2$ region around the focus. We record and average the polarimetric properties over a $10\times10$ grid, covering an area of $250\times250\,\text{nm}^2$ around the estimated center, which is indicated by reduced transmission in \figref{fig:fig2abc}$\boldsymbol{b}$.\\
	To reconstruct the $4\times 4$ Mueller matrix $\mathbf{M}$, determining the polarization response of the system, we must invert the following identity~\cite{arteaga_o._mueller_2010}:
\begin{align}	
\mathbf{S}_o = \mathbf{M} \mathbf{S}_i
\text{,}
\label{eq:eq1} \end{align}
where $\mathbf{S}_i$ and $\mathbf{S}_o$ are the input and the output Stokes vectors, respectively. To this end, we illuminate our sample with an overdetermined set of six input polarization states, estimated to be $\left\{\text{RCP, LCP, X, Y, diagonal, antidiagonal} \right\}$, and analyze the transmitted light. For each position of the sample in the grid $(x,y)$, we acquire angularly resolved output Stokes vectors $\hat{\mathbf{S}}_o{(x,y,\mathbf{k})}$ and integrate them $\mathbf{S}_o(x,y)= \iint \hat{\mathbf{S}}_o{(x,y,\mathbf{k})} \mathrm{d}^2\mathbf{k}$ over the angular region of $\text{NA}_{\text{eff}}= 0.4$, where we detect the far-field interference of incident and scattered light (\figref{fig:fig2abc}$\boldsymbol{c}$). Finally, averaging $\mathbf{S}_o(x,y)$ over the grid provides us with the desired output Stokes vectors $\mathbf{S}_o = \overline{ \mathbf{S}_o(x,y)} $. We also determine the actual experimental input Stokes vectors $\mathbf{S}_i$ by performing the same procedure on a plain glass substrate. Using these six input and six output polarizations, we invert Eq.~\ref{eq:eq1} and obtain the experimental Mueller matrix $ \mathbf{M}_{exp}$. However, experimental noise, the finite integration region of $\text{NA}_{\text{eff}}= 0.4$ and averaging of the Stokes vectors over the grid may result in a matrix $\mathbf{M}_{exp}$, which is unphysical and contains depolarization, inhibiting the analysis of CB and CD. Therefore, we apply Cloude's sum decomposition \cite{cloude_conditions_1990}, providing us with the closest physical and non-depolarizing estimate $\mathbf{M}$ of $\mathbf{M}_{exp}$. Finally, we calculate CB and CD from the elements $m_{ij}$ of $\mathbf{M}$~\cite{arteaga_o._mueller_2010}:
\begin{align} \begin{split}
\text{CB}&=0.5[m_{12}-m_{21}]\text{,}\\
\text{CD}&=0.5[m_{03}+m_{30}]
\text{.}
\end{split} \label{eq:eq2} \end{align}
In \figref{fig:fig3abcd}$\boldsymbol{a}$ and \ref{fig:fig3abcd}$\boldsymbol{c}$, we present the obtained experimental results for CB and CD, respectively, quantifying the chiroptical response. For comparison, \figref{fig:fig3abcd}$\boldsymbol{b}$ and \ref{fig:fig3abcd}$\boldsymbol{d}$ show results obtained from FDTD simulations. The nanotrimer is modeled as a system of three perfect spheres with radii $r=90\,$nm and inter-particle gaps of $2\,$nm, placed on a glass substrate ($n=1.52$) and arranged in the geometry shown in \figref{fig:fig1}$\boldsymbol{a}$. The Si nanospheres are surrounded by a $\text{SiO}_2$~\cite{edited_by_edward_d._palik_handbook_1985} shell with estimated thickness of $8\,$nm, correspondingly reducing the core diameter.\\
	The actual optical handedness of the heterogeneous trimer on substrate in \figref{fig:fig3abcd} depends on the wavelength and changes sign in the investigated spectral range of $500\,\text{nm}\le\lambda\le700\,$nm. Moreover, for an isolated nanostructure, the chiroptical response is exceptionally strong. The experimentally measured CB corresponds to a maximum optical rotation of about \ang{-0.8}. Assuming a thickness of $180\,$nm for the sample, this corresponds to a refractive index difference for LCP and RCP of $|\Delta n| \approx 0.015$, which is an extremely high value as compared to natural optically active media (typically $|\Delta n| \sim 10^{-5}$). Remarkably, the spectra show a characteristic fingerprint of the Born-Kuhn model dispersion \cite{yin2013_interpreting}, manifested by the prominent dip in CB, accompanied by a zero-crossing for the bisignate CD, which appear around $\lambda \approx 580\,$nm in \figref{fig:fig3abcd}$\boldsymbol{a}$, \ref{fig:fig3abcd}$\boldsymbol{c}$ and $\lambda \approx 600\,$nm in \figref{fig:fig3abcd}$\boldsymbol{b}$, \ref{fig:fig3abcd}$\boldsymbol{d}$.
			\\
	Qualitatively, we achieve a good overlap between the numerically and experimentally retrieved spectra. The blue-shift of the CB and CD spectra with respect to FDTD simulations may be attributed to underestimation of the SiO$_2$ shell thickness. Additionally, the deviations of the actual sample from the ideally assumed geometry, clearly visible under the scanning electron microscope in \figref{fig:fig1}$\boldsymbol{b}$, are not accounted for in simulations. As discussed earlier, any asymmetry affects the circular anisotropies, reasoning the observation of substantially higher CB and CD in experiment. Most importantly, our nanosphere assembly is extremely sensitive to the inter-particle gaps, which cannot be determined exactly and which considerably influence the optical response \cite{albella_low-loss_2013}. Lastly, the individual heterogeneous nanotrimer is strongly anisotropic and exhibits linear birefringence (LB) and linear dichroism (LD), which are orders of magnitude larger than CB and CD. Strong LB and LD are known to induce artifacts in measurements of CB and CD~\cite{oriented_mol_1987,arteaga_o._mueller_2010}. In the supplemental material~\cite{supp}, we compare the experimentally reconstructed and the simulated spectra of LB and LD. Additionally, we numerically investigate forward and backward illumination of the sample, a nanotrimer in a homogeneous environment and a nanotrimer of opposite handedness~\cite{supp}. 
		%
%
	\section{Discussion and Conclusion}
	Owing to the relatively high chiral response of an individual nanostructure, such heterogeneous systems hold promise for constructing flat chiral and on-chip optical elements. First, the chiral response may be significantly enhanced by utilizing a higher refractive index substrate, by introducing structural chirality, by tailoring individual resonances of the constituents and by arranging the nanostructures in arrays which support lattice resonances. Secondly, owing to the heterogeneous environment, each of the nanoparticles in the nanotrimer assembly responds differently to the incident field~\cite{banzer_chiral_2016}. The latter suggests that such systems can potentially ``sense'' the gradient of the excitation field and distinguish the topological charge of orbital angular momentum of incident beams~\cite{PhysRevB.99.075155,wozniak_interaction_2019}, paving the way towards novel nanoscopic sensors and sorters.\\
In conclusion, we have experimentally investigated a geometrically symmetric heterogeneous nanotrimer on a glass substrate. The in-plane reflection symmetry of the system is broken by its heterogeneous material composition, while the glass substrate breaks the mirror symmetry of the whole system, transitioning its morphology from achiral to chiral. We have experimentally reconstructed the circular birefringence and circular dichroism spectra. The study of an individual nanostructure allowed us to preclude any contributions of delocalized lattice effects, collective resonances and near-field coupling effects, otherwise present in arrays of nanostructures. Additionally, our study provides a clear-cut distinction between the material- and geometry-induced chiroptical responses in a system exhibiting a heterogeneity- and substrate-induced break of symmetry, shedding light on chiral light-matter interactions at substrate-nanostructure interfaces.\\
	%

	\begin{acknowledgments}
		The authors gratefully acknowledge fruitful discussions with Israel De Leon, Gerd Leuchs and Pawe{\l} Wo{\'z}niak.
	\end{acknowledgments}

	\bibliographystyle{apsrev4-1}
	\bibliography{trimer_bib}
	
\end{document}


\def\theequation{S\arabic{equation}}
\def\thefigure{S\arabic{figure}}

\title{Substrate-Induced Chirality in an Individual Nanostructure:\\Supplemental Material}

\author{Sergey Nechayev} 
	\thanks{These authors contributed equally}
	 \affiliation{Max Planck Institute for the Science of Light, Staudtstr. 2, D-91058 Erlangen, Germany}
	 \affiliation{Institute of Optics, Information and Photonics, University Erlangen-Nuremberg, Staudtstr. 7/B2, D-91058 Erlangen, Germany}

\author{Ren\'{e} Barczyk}
  \thanks{These authors contributed equally}
	\affiliation{Max Planck Institute for the Science of Light, Staudtstr. 2, D-91058 Erlangen, Germany}
	\affiliation{Institute of Optics, Information and Photonics, University Erlangen-Nuremberg, Staudtstr. 7/B2, D-91058 Erlangen, Germany}


\author{Uwe Mick}
	\affiliation{Max Planck Institute for the Science of Light, Staudtstr. 2, D-91058 Erlangen, Germany}
	\affiliation{Institute of Optics, Information and Photonics, University Erlangen-Nuremberg, Staudtstr. 7/B2, D-91058 Erlangen, Germany}

\author{Peter Banzer}
\email{peter.banzer@mpl.mpg.de}
\homepage{http://www.mpl.mpg.de/}
	\affiliation{Max Planck Institute for the Science of Light, Staudtstr. 2, D-91058 Erlangen, Germany}
	\affiliation{Institute of Optics, Information and Photonics, University Erlangen-Nuremberg, Staudtstr. 7/B2, D-91058 Erlangen, Germany}
\date{\today}

\maketitle
\section*{Glossary}
\noindent  $\mathbf{M}$ --- Mueller matrix\\
$m_{ij}$ --- element ($i,j = 0\dots3$) of $\mathbf{M}$\\
CB --- circular birefringence\\
CD --- circular dichroism\\
LB --- linear birefringence\\
LD --- linear dichroism\\
CPL --- circularly polarized light\\
$h$ --- horizontal\\
$d$ --- diagonal\\
RCP --- right-hand circular polarization\\
LCP --- left-hand circular polarization\\
QPN --- quasi-planar nanostructure\\ 
\section{Supplementary Note 1:\\Linear anisotropies of the nanotrimer}
\begin{figure*}[!htb]
  \includegraphics[width=0.99\textwidth]{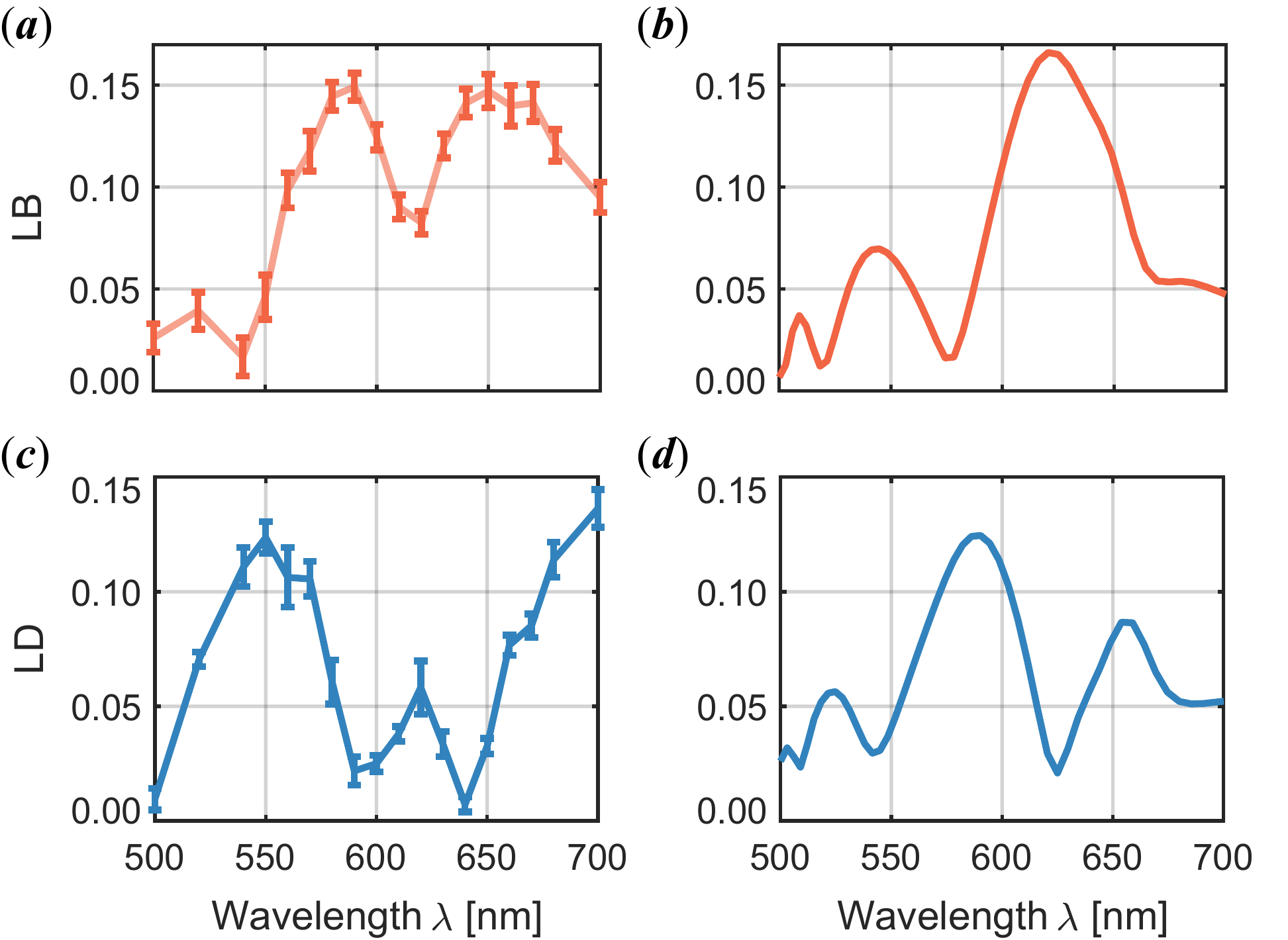}
  \caption{Experimentally measured ($\boldsymbol{a},\,\boldsymbol{c}$) and numerically calculated ($\boldsymbol{b},\,\boldsymbol{d}$) spectra of maximum linear birefringence (LB) and maximum linear dichroism (LD).}
  \label{fig:figs1}
\end{figure*}

The complete Mueller matrix k-space spectropolarimetry that we experimentally perform on our heterogeneous nanotrimer on a glass substrate allows us to also estimate the linear anisotropies. Linear anisotropies, however, must be defined with respect to a specific axis and therefore, contrary to circular anisotropies, do not have cylindrical symmetry. To obtain practical cylindrically symmetric parameters, representing linear birefringence (LB) and linear dichroism (LD), we first obtain the projections of LB and LD onto the horizontal ($LB_h = 0.5[m_{32} - m_{23}]$, $LD_h = -0.5[m_{10} + m_{01}]$) and {45\textdegree} diagonal ($LB_d = 0.5[m_{13} - m_{31}]$, $LD_d = -0.5[m_{20} + m_{02}]$) laboratory axes from the elements $m_{ij}$ ($i,j = 0\dots3$) of the reconstructed Mueller matrix $\mathbf{M}$. Is is important to note that, due to their definition, $LB_h$, $LD_h$, $LB_d$ and $LD_d$ depend on the actual orientation of the sample with respect to the arbitrarily chosen laboratory axes. In order to avoid alignment uncertainties when comparing experimental values to numerical predictions~\cite{noauthor_fdtd_nodate}, we calculate the maximum LB and LD in the transverse plane~\cite{arteaga_o._mueller_2010}, defined as LB$=\sqrt{LB_h^2+LB_d^2}$ and LD$=\sqrt{LD_h^2+LD_d^2}$. Maximum LB and LD are invariant to rotation of the sample in the transverse plane (around the $z$-axis). Moreover, in most cases, the principal axes of LB and LD, i.e. the directions in a material along which these quantities are maximal, coincide. In Fig.~\ref{fig:figs1}, we compare the experimentally reconstructed to the numerically obtained values of LB and LD~\cite{noauthor_fdtd_nodate}. The linear anisotropies exhibit several peaks of magnitude $\sim 0.15$ in the investigated spectral range. Taking the sphere diameter of $d=180\,$nm as thickness of the sample and utilizing the definition of LB for natural materials LB$=\frac{2 \pi d}{\lambda} \Delta n$, leads to a maximal refractive index difference of $|\Delta n|\approx 0.078$ at $\lambda = 590\,$nm. This is a fairly high value, comparable to the birefringence commonly found in uniaxial crystals such as lithium niobate (LiNbO$_3$, $|\Delta n| = 0.085$ at $\lambda = 590\,$nm). Similarly to circular birefringence (CB) and circular dichroism (CD), the experimental values of LB and LD are blue-shifted with respect to their numerical counterparts due to the sample oxidation.
\section{Supplementary Note 2:\\Comparison of forward and backward illumination}
\begin{figure*}[!htb]
  \includegraphics[width=0.99\textwidth]{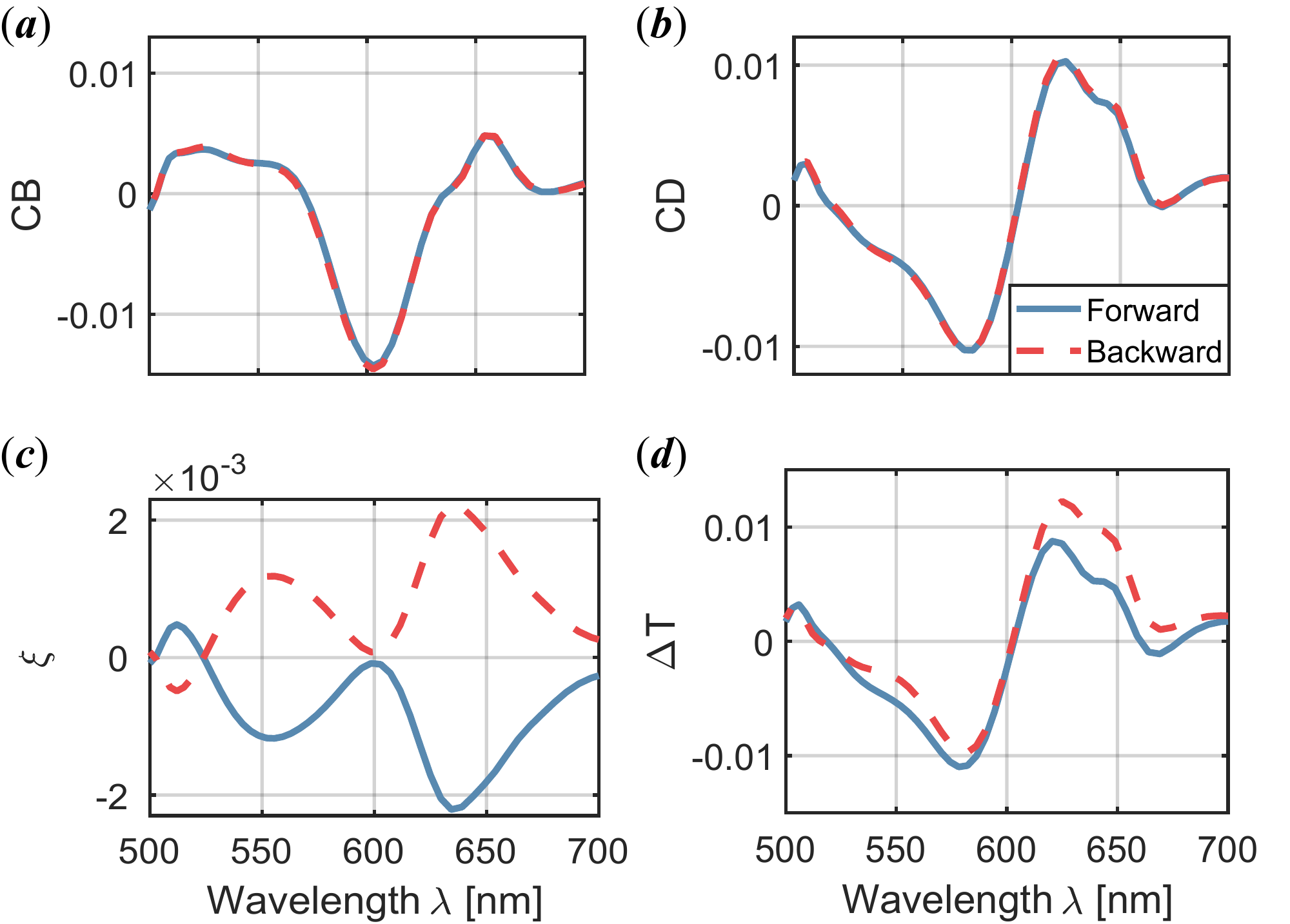}
  \caption{Comparison of the optical response for illumination of the nanotrimer from the air side (forward, blue) and the substrate side (backward, red). ($\boldsymbol{a}$) Circular birefringence (CB) and ($\boldsymbol{b}$) circular dichroism (CD) are exactly the same for a forward and a backward propagating wave, underlining the reciprocity of the chiroptical response. ($\boldsymbol{c}$) In contrast, the linear anisotropy parameter $\xi$, responsible for the asymmetric conversion of circularly polarized light (CPL), is inverted upon wavevector reversal. ($\boldsymbol{d}$) The interplay of wavevector-invariant CB and CD and inverted linear anisotropies leads to an asymmetric response in the differential transmission $\Delta T$ of CPL.}
  \label{fig:figs2}
\end{figure*}

The strong linear anisotropies of the nanotrimer indicate a pronounced asymmetric response to circularly polarized light (CPL), irrespective of additionally present circular anisotropies. For instance, the linear anisotropy parameter $\xi =	0.5[LB_h\, LD_d - LB_d\,LD_h] = 0.5[m_{03} - m_{30}]$ is responsible for the asymmetry of CPL conversion from left-hand circular polarization (LCP) to right-hand circular polarization (RCP) and vice versa~\cite{arteaga_o._mueller_2010}, which may be confused with a true chiroptical response. However, as discussed earlier, linear anisotropies invert their sign upon wavevector reversal, while circular anisotropies preserve it. Fig.~\ref{fig:figs2} numerically compares forward and backward illumination of the nanotrimer on a substrate. Fig.~\ref{fig:figs2}$\boldsymbol{a}$ and~\ref{fig:figs2}$\boldsymbol{b}$ show that CB and CD do note change with the reversal of the illumination direction. At the same time, as shown in Fig.~\ref{fig:figs2}$\boldsymbol{c}$, $\xi$ flips its sign upon wavevector reversal. The interplay of wavevector-invariant circular anisotropies and inverted linear anisotropies leads to an asymmetric differential transmission of CPL $\Delta T= 0.5[T_{LCP} - T_{RCP}]= m_{03}=CD+\xi$, as can be seen in Fig.~\ref{fig:figs2}$\boldsymbol{d}$. It must be noted that the average value of $\Delta T$ for forward and backward illumination $\left\langle  \Delta T  \right\rangle = 0.5 [CD+\xi ] + 0.5 [CD-\xi ] $ represents CD when properly recorded along the illumination direction, avoiding the scattered light contribution.
\section{Supplementary Note 3:\\Comparison of nanotrimers in free-space and on a substrate}
\begin{figure*}[!htb]
  \includegraphics[width=0.99\textwidth]{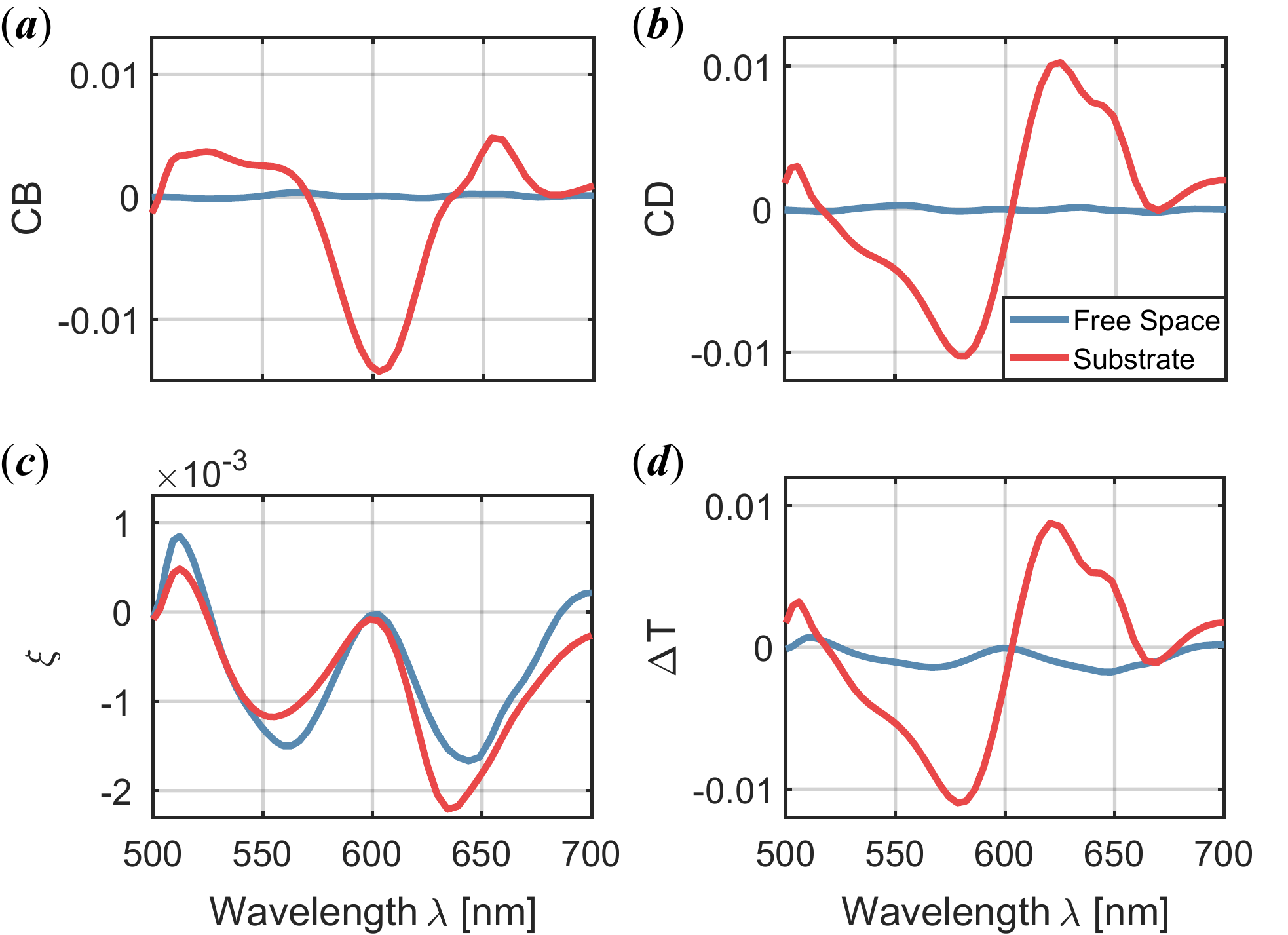}
  \caption{Comparison of the optical response of the nanotrimer in free-space and on a substrate. ($\boldsymbol{a}$) Circular birefringence (CB) and ($\boldsymbol{b}$) circular dichroism (CD) spectra serve as an indicator for true chirality and vanish for the nanotrimer in free-space. ($\boldsymbol{c}$) In contrast, the linear anisotropy parameter $\xi$, responsible for the asymmetric conversion of circularly polarized light (CPL), is non-zero in both cases. ($\boldsymbol{d}$) Non-zero linear anisotropies lead to non-zero differential transmission $\Delta T$ of CPL also in free-space.}
  \label{fig:figs3}
\end{figure*}

The sample we measure experimentally is indeed chiral, which is optically manifested by CB and CD emerging at normal incidence. Albeit the lack of in-plane reflection symmetry, the sample plane represents a mirror plane of  quasi-planar nanostructures (QPNs), rendering the heterogeneous nanotrimer itself structurally achiral. It is only the presence of a substrate which breaks the mirror symmetry of the system as a whole and results in a transition of the morphology from achiral to chiral. CD and CB would not be present for a nanotrimer in a homogeneous environment, since true reciprocal chiroptical effects are strictly forbidden for QPNs illuminated at normal incidence due to symmetry and reciprocity constraints~\cite{kuwata-gonokami_giant_2005,Maslovski_2009,arteaga_o._mueller_2010,arteaga_relation_2016}.\\
In Fig.~\ref{fig:figs3}, we compare CB and CD spectra calculated for the nanotrimer with and without substrate. As anticipated, CB and CD vanish (up to negligible numerical instabilities) for the nanotrimer in a homogeneous environment, identifying the substrate-induced interaction as the origin of the reciprocal chiroptical response. In contrast, the linear anisotropy parameter $\xi$ shown in Fig.~\ref{fig:figs2}$\boldsymbol{c}$, responsible for the asymmetry of CPL conversion, is non-zero also in free-space, as appears in Fig.~\ref{fig:figs3}$\boldsymbol{c}$. As discussed earlier, $\xi \neq 0$ leads to an asymmetric differential transmission $\Delta T \neq 0$ of CPL also in free-space, as can be seen in Fig.~\ref{fig:figs3}$\boldsymbol{d}$. However, $\Delta T$ should not be confused with the chiroptical response manifested by CB and CD. The full Mueller matrix k-space spectropolarimetry of the transmitted light is capable of separating and quantifying the individual linear and circular contributions to the optical response and reveals the fundamental difference between linear anisotropies and the true chiroptical response.~\cite{arteaga_o._mueller_2010,arteaga_relation_2016}.
\section{Supplementary Note 4:\\Comparison of nanotrimers of opposite handedness}
\begin{figure*}[!htb]
  \includegraphics[width=0.99\textwidth]{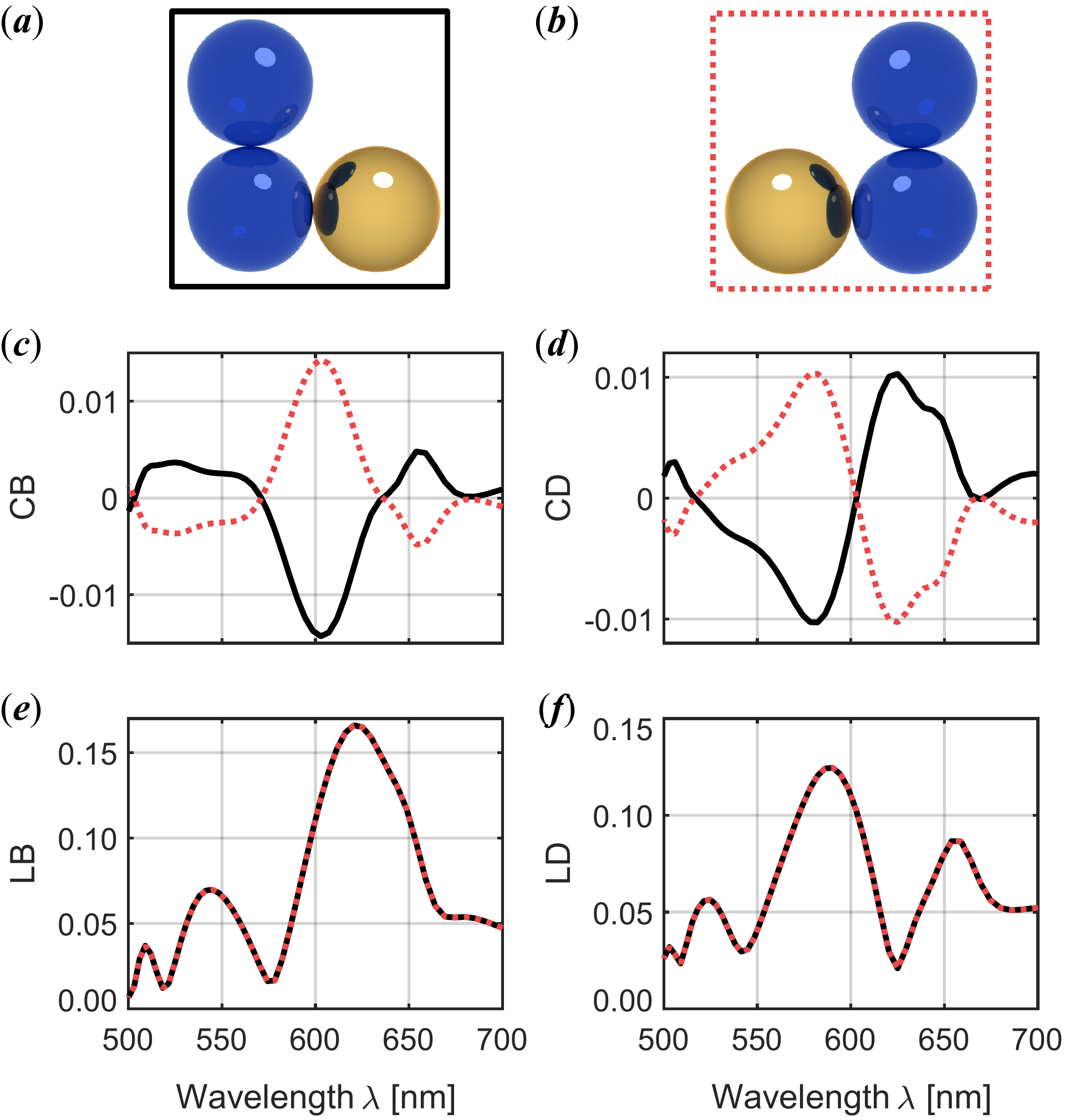}
  \caption{Optical response of oppositely handed nanotrimers on a substrate. The solid black and dotted red curves correspond to the geometry of ($\boldsymbol{a}$) the experimentally investigated sample and ($\boldsymbol{b}$) the oppositely handed trimer, respectively. The spectra of ($\boldsymbol{c}$) circular birefringence (CB) and ($\boldsymbol{d}$) circular dichroism (CD), quantifying the chiroptical response, are inverted for oppositely handed trimers. Contrarily, the spectra of ($\boldsymbol{e}$) maximum linear birefringence (LB) and ($\boldsymbol{f}$) maximum linear dichroism (LD) are exactly the same in both cases.}
  \label{fig:figs4}
\end{figure*}

Contrary to linear anisotropies, the spectra of the true chiroptical properties CD and CB are only inverted when changing the handedness of the nanotrimer. The numerical plots of the CB and CD spectra for two oppositely handed nanotrimers on a substrate (Fig.~\ref{fig:figs4}$\boldsymbol{a}$ and~\ref{fig:figs4}$\boldsymbol{b}$) are shown in Fig.~\ref{fig:figs4}$\boldsymbol{c}$ and~\ref{fig:figs4}$\boldsymbol{d}$, respectively. As expected, CB and CD have exactly the same magnitude but opposite sign for oppositely handed enantiomers, whereas LB and LD are the same for both nanotrimers, as shown in Fig.~\ref{fig:figs4}$\boldsymbol{e}$ and~\ref{fig:figs4}$\boldsymbol{f}$, respectively~\cite{arteaga_o._mueller_2010,arteaga_relation_2016}. Moreover, LB and LD are approximately one order of magnitude higher than CB and CD, which is known to have a disturbing influence on the experimental measurement of the circular anisotropies~\cite{arteaga_o._mueller_2010,arteaga_relation_2016}. These considerations underline the necessity of the utilized Mueller matrix k-space spectropolarimetry characterization technique, capable of separating and properly quantifying all contributions to the polarimetric response~\cite{arteaga_o._mueller_2010,arteaga_relation_2016}.

\pagebreak

\pagebreak

\pagebreak

\bibliography{trimer_bibs}